# Lagrange Model for the Chiral Optical Properties of Stereometamaterials

H. Liu[1], J. X. Cao[1], S. N. Zhu[1]

[1]*Department of Physics, National Laboratory of Solid State Microstructures, Nanjing University, Nanjing 210093 People's Republic of China*
http://dsl.nju.edu.cn/mpp

N. Liu[2], R. Ameling[2], and H. Giessen[2]

[2]*4th Physics Institute, University of Stuttgart, Stuttgart, Germany*
http://www.pi4.uni-stuttgart.de/

We employ a general Lagrange model to describe the chiral optical properties of stereometamaterials. We derive the elliptical eigenstates of a twisted stacked split-ring resonator, taking phase retardation into account. Through this approach, we obtain a powerful Jones matrix formalism which can be used to calculate the polarization rotation, ellipticity, and circular dichroism of transmitted waves through stereometamaterials at any incident polarization. Our experimental measurements agree well with our model.

Stereo in Greek means spatial or three-dimensional. Stereometamaterials are three-dimensional nanostructures with characteristic dimensions much smaller than the wavelength of light [1]. Examples of stereometamaterials include stacked wires [2], crosses [3], gammadions [4], and spirals [5]. One specific type of stereometamaterials are stacked and twisted split-ring resonators (SRR) [6-9]. In particular, upon twisting of the two individual split-rings at angles other than 0 and 180 degrees, such a stereometamaterial can form left- or right handed enantiomers, which might exhibit chiral optical properties. In analogy to stereochemistry [10], the transmission and reflection properties of stereometamaterials can be measured with linearly and circularly polarized light at different wavelengths, and quantities such as circular dichroism, polarization rotation, and ellipticity can be determined experimentally.

A convenient description of the polarization properties of optical elements at normal incidence is the Jones calculus, which connects the input and output polarization vector of light by a 2x2 matrix. The polarization eigenstates are the eigenvectors of that Jones matrix. In order to design future polarization elements composed of nanooptical components, one requires a simple coupling model which allows for intuitive understanding, as well as for the calculation and application of the polarization properties in nanophotonic stereometamaterials.

In this Letter, we provide such a model. Namely, we introduce a Lagrange model that takes the individual oscillators of our stereometamaterial plus important coupling processes into account. In particular, we treat electric and magnetic coupling separately. Importantly, we include phase retardation, which is essential for stacked structures. It arises from the light propagation along the stacked individual elements as well as from the resonant behavior of the SRR elements. We are able to determine the chiral optical eigenmodes. This allows us to describe even very complex structures such as multi-layer twisted structures with large spacer distances as described in [11]. For a planar structure without the phase retardation effect, the two eigenwaves will degrade to linear polarization and optical activity at normal incidence will disappear. Our model will allow tuning of the polarization properties in complex coupled 3D systems, such as nanospirals, to a set of desired parameters and enable for example broadband operation of such spirals as near-achromatic wave retarders [12].

In principle, ab-initio methods such as FDTD, S-Matrix [13], and others, are able to predict the polarization properties of stereometamaterials. However, these methods solve Maxwell's equations numerically and do not provide deeper insight into the intricate coupling mechanisms.

Figure 1 displays the geometry of our stereometamaterial together with the design parameters. Each unit cell is composed of two separate SRRs, which are twisted at by an angle $\theta$ with respect to one another. The twist angle varies from 0 to 180°. The SRRs are embedded in a homogeneous dielectric with n=1.55 (our spacer material PC403) and reside on a glass substrate (n=1.51). The incident EM wave is normal to the SRRs and propagates in the z-direction as shown in Fig.1. To study the EM response of the stereo-SRR metamaterial, we rely on a full wave simulation using a time-domain simulation software (CST Microwave Studio). In the simulation, the metal permittivity is given by $\varepsilon(\omega) = 1 - \omega_p^2 / (\omega^2 + i\omega_\tau \omega)$, where $\omega_p$ is the bulk plasma frequency and $\omega_p$ is the relaxation rate. For gold, the characteristic frequencies fitted to experimental data are $\omega_p = 1.37 \times 10^4 THz$ and $\omega_\tau = 40.84 THz$ [14].

A single SRR consists of a magnetic loop (the metal ring) with inductance $L$ and a capacitor with capacitance $C$ (corresponding to the slit). Thus it can be modeled by an ideal L-C-circuit with resonance frequency $\omega_0 = 1/\sqrt{LC}$, for which the Lagrangian can be written as $\mathcal{L}_0 = L(\dot{Q}^2 - \omega_0^2 Q^2)/2$. Here the total oscillatory charge $Q$ accumulated in the gap is defined as a generalized coordinate, $L\dot{Q}^2/2$ is the kinetic energy of the oscillations, and $L\omega_0^2 Q^2/2$ refers to the electrostatic energy stored in the gap. In the hybridization model, each SRR can be regarded as an artificial atom; while the stereo-SRR system represents an artificial molecule. When excited by an incident EM wave, the Lagrangian of the stereo-SRR metamaterial can be rewritten as:

$$\mathcal{L} = L(\dot{Q}_1^2 - \omega_0^2 Q_1^2)/2 + L(\dot{Q}_2^2 - \omega_0^2 Q_2^2)/2 + M_m \dot{Q}_1 \dot{Q}_2 \\ - M_e \omega_0^2 Q_1 Q_2 \cdot (\cos\theta - \alpha \cdot \cos^2\theta + \beta \cdot \cos^3\theta) - \mathbf{P}_1 \cdot \mathbf{E} - \mathbf{P}_2 \cdot \mathbf{E} \cdot e^{i\varphi} \quad (1)$$

The first two terms of Eq.(1) give the energy stored in the two SRRs, while the last two terms are the magnetic and electric interaction energy between them. The contributions of the electric quadrupolar and octupolar interactions are considered as correction terms of the electric dipolar action, with coefficients α and β, respectively. $\mathbf{P}_1 = l_{eff} \cdot \hat{\mathbf{x}} \cdot Q_1$ and $\mathbf{P}_2 = l_{eff} \cdot (\cos\theta \cdot \hat{\mathbf{x}} - \sin\theta \cdot \hat{\mathbf{y}}) \cdot Q_2$ are the induced electric dipole moments in the gap of two SRRs ($l_{eff}$ is the effective length of the single SRR), $\varphi$ is the phase retardation which arises due to the distance between two stacked SRRs in the propagation direction and due to the resonant behavior of the SRR elements. According to reference [6], the two eigenfrequencies of the stereo-SRR metamaterial are obtained as: $\omega_- = \omega_0 \cdot \sqrt{(1+\kappa_e)/(1+\kappa_m)}$ and $\omega_+ = \omega_0 \cdot \sqrt{(1-\kappa_e)/(1-\kappa_m)}$, where $\kappa_m = M_m/L$ and $\kappa_e = M_e \cdot (\cos\theta - \alpha \cdot \cos^2\theta + \beta \cdot \cos^3\theta)/L$ are the magnetic and electric coupling coefficients, respectively. The electric interaction of the stereo-SRR metamaterial is closely related to the spatial arrangement of the metamaterial elements. Fitting the twist dispersion curves yields the corresponding coefficients as $\kappa_m = M_m/L = 0.09$, $\kappa_e' = M_e/L = 0.14$, $\alpha = 0.8$ and $\beta = -0.4$.

By introducing the Ohmic dissipation $\mathfrak{R} = \sigma(\dot{Q}_1^2 + \dot{Q}_2^2)/2$ and substituting it into the Euler-Lagrange equations

$$(d/dt)(\partial\mathcal{L}/\partial\dot{Q}_i) - \partial\mathcal{L}/\partial Q_i = -\partial\mathfrak{R}/\partial\dot{Q}_i \quad (i=1,2) \quad (2)$$

we obtain the solutions for $Q_i$ ($i=1,2$) [15]. Based on the material equations $\mathbf{P} = \mathbf{P}_1 + \mathbf{P}_2 = \chi\mathbf{E} = (1-\varepsilon)\mathbf{E}$ and combining them with the definition of $\mathbf{P}$ in Eq. (1), the elements of the effective permittivity tensor for the stereo-SRR metamaterial can be obtained as in the reference [15]. According to the wave equation $\mathbf{k} \times (\mathbf{k} \times \mathbf{E}) + \omega^2 \varepsilon \mu \cdot \mathbf{E} = 0$, the refractive indices of the two eigenmodes of such a stereo-SRR metamaterial are given as:

$$n_\pm = \sqrt{\left(\varepsilon_{xx} + \varepsilon_{yy} \pm \sqrt{(\varepsilon_{xx} - \varepsilon_{yy})^2 + 4\varepsilon_{xy}\varepsilon_{yx}}\right)/2} \quad (3)$$

The corresponding polarization states of these two eigenmodes are represented by the following Jones vectors:

$$\mathbf{J}_- = \frac{1}{\mathcal{N}_-} \begin{pmatrix} -2\varepsilon_{yx} \\ (\varepsilon_{xx} - \varepsilon_{yy}) + \sqrt{(\varepsilon_{xx} - \varepsilon_{yy})^2 + 4\varepsilon_{xy}\varepsilon_{yx}} \end{pmatrix}$$
$$\mathbf{J}_+ = \frac{1}{\mathcal{N}_+} \begin{pmatrix} -2\varepsilon_{xy} \\ (\varepsilon_{xx} - \varepsilon_{yy}) - \sqrt{(\varepsilon_{xx} - \varepsilon_{yy})^2 + 4\varepsilon_{xy}\varepsilon_{yx}} \end{pmatrix} \quad (4)$$

where $\mathcal{N}_\pm$ are the corresponding normalized coefficients. Similar to E. U. Condon's theory [16], these two Jones vectors present two orthogonal elliptically polarized states $\mathbf{J}_- \cdot \mathbf{J}_+ = 0$. For these two eigenmodes, the corresponding transmission coefficients through the stereometamaterial can be obtained as [17]

$$t_\pm = \frac{4n_\pm \exp(-in_\pm kd)}{(n_\pm + 1)^2 - (n_\pm - 1)^2 \exp(-2in_\pm kd)} \quad (5)$$

Any state in the x-y coordinate system can be transformed into the principal $\mathbf{J}_+$-$\mathbf{J}_-$ coordinate system. Therefore, the transmission coefficient for any incident wave $\begin{pmatrix} E_x^i \\ E_y^i \end{pmatrix}$ can be calculated as

$$\begin{pmatrix} E_x^t \\ E_y^t \end{pmatrix} = \mathbf{T} \cdot \begin{pmatrix} E_x^i \\ E_y^i \end{pmatrix} = \begin{pmatrix} \hat{x} \cdot \mathbf{J}_+ & \hat{x} \cdot \mathbf{J}_- \\ \hat{y} \cdot \mathbf{J}_+ & \hat{y} \cdot \mathbf{J}_- \end{pmatrix} \cdot \begin{pmatrix} t_+ & 0 \\ 0 & t_- \end{pmatrix} \cdot \begin{pmatrix} \hat{x} \cdot \mathbf{J}_+ & \hat{y} \cdot \mathbf{J}_+ \\ \hat{x} \cdot \mathbf{J}_- & \hat{y} \cdot \mathbf{J}_- \end{pmatrix} \cdot \begin{pmatrix} E_x^i \\ E_y^i \end{pmatrix} \quad (6)$$

Here, $\mathbf{T}$ is the Jones matrix which is used to determine the polarization state of the transmission wave through the stereometamaterial. Eq.(6) is a very powerful result. In principle, based on this equation, we can calculate not only the transmission of the stereometamaterials but also the polarization states of transmitted waves with different incident polarization for any frequency. Our stereometamaterial can therefore be replaced by a black box which describes an artificial polarization element with a certain Jones Matrix $\mathbf{J}_+$ and $\mathbf{J}_-$ [15].

For the linearly polarized incident case $\begin{pmatrix} 1 \\ 0 \end{pmatrix}$, the transmitted wave is $\begin{pmatrix} t_- \hat{x} \cdot \mathbf{J}_+ \hat{x} \cdot \mathbf{J}_+ + t_- \hat{x} \cdot \mathbf{J}_- \hat{x} \cdot \mathbf{J}_- \\ t_+ \hat{y} \cdot \mathbf{J}_+ \hat{x} \cdot \mathbf{J}_+ + t_- \hat{y} \cdot \mathbf{J}_- \hat{x} \cdot \mathbf{J}_- \end{pmatrix}$. The polarization state of the transmitted wave will change accordingly. For any other polarized wave, such as left-handed and right-handed circularly polarized waves (LCP and RCP), the polarization state of the transmission wave can also be calculated directly based on the above Jones Matrix model.

For stereo-SRR structures with different twist angles $\theta$, the optical response is quite different. For the cases $\theta = 0°$ and $180°$, $\varepsilon_{xy} = \varepsilon_{yx} = 0$, $\varepsilon_{yy} = 1$, $\varepsilon_{xx} \neq 1$, only the x-polarization resonance can be excited. Correspondingly, $n_+ = \sqrt{\varepsilon_{xx}}$, $n_- = 1$, $\mathbf{J}_+ = \begin{pmatrix} 1 \\ 0 \end{pmatrix}$, $\mathbf{J}_- = \begin{pmatrix} 0 \\ 1 \end{pmatrix}$. For the case $\theta = 90°$, the refractive indices of the two eigenmodes and the corresponding eigenwavevectors are calculated as

$$n_\pm = \sqrt{1 + \frac{(1+e^{i\varphi})\ell(\omega^2 - i\gamma\omega - \omega_0^2) \pm \ell\sqrt{(1-e^{i\varphi})^2(\omega^2 - i\gamma\omega - \omega_0^2)^2 + 4\kappa_m^2\omega^4 e^{i\varphi}}}{2(1-\kappa_m^2) \cdot (\omega^2 - i\gamma_-\omega - \omega_-^2) \cdot (\omega^2 - i\gamma_+\omega - \omega_+^2)}} \quad (7)$$

$$\mathbf{J}_{-} = \frac{1}{N_{-}} \begin{pmatrix} -2\kappa_m \omega^2 \\ (1-e^{i\varphi})(\omega^2 - i\gamma\omega - \omega_0^2) + \sqrt{(1-e^{i\varphi})^2(\omega^2 - i\gamma\omega - \omega_0^2)^2 + 4\kappa_m^2 \omega^4 e^{i\varphi}} \end{pmatrix}$$

$$\mathbf{J}_{+} = \frac{1}{N_{+}} \begin{pmatrix} -2\kappa_m \omega^2 e^{i\varphi} \\ (1-e^{i\varphi})(\omega^2 - i\gamma\omega - \omega_0^2) - \sqrt{(1-e^{i\varphi})^2(\omega^2 - i\gamma\omega - \omega_0^2)^2 + 4\kappa_m^2 \omega^4 e^{i\varphi}} \end{pmatrix}$$ (8)

Here, $\ell = l_{eff}\tau = l_{eff}^2 / L$, $\gamma = \sigma/L$, $\gamma_{-} = \gamma/(1+\kappa_m)$ and $\gamma_{+} = \gamma/(1-\kappa_m)$. In order visualize the above analytical results, we plot the $n_{\pm}$ in Fig.2 (a). Eq.(8) implies that we have varying eigenvectors $\mathbf{J}_{\pm}$ with varying polarization state at different incident frequencies. For example, at 140 THz, the two eigenmodes are ellipses as plotted in Fig.2 (d).

For a planar structure, there is no phase retardation effect between two SRRs. The phase retardation $\varphi$ equals zero, and hence the two eigenmodes $\mathbf{J}_{\pm}$ are linearly polarized. For the stereo-SRR structure, the calculated value for $\varphi$ is displayed in Fig. 2(c). Obviously, $\varphi$ is not equal to zero around the resonance frequencies and hence the eigenstates are not linearly polarized. *Therefore, the phase retardation $\varphi$ is the decisive parameter that distinguishes stereometamaterials from planar metamaterials with respect to their chiral optical properties.* Furthermore, we can change the phase retardation $\varphi$ by changing the coupling distance. As a result, the eigenmodes of stereometamaterials and their propagation properties can also be tuned this way.

For any other incident polarized wave $\mathbf{E} = \begin{pmatrix} E_x \\ E_y \end{pmatrix}$, which can be regarded as the combination of two eigenwaves $\begin{pmatrix} E_x \\ E_y \end{pmatrix} = a_{-}\mathbf{J}_{-} + a_{+}\mathbf{J}_{+}$, two eigenmodes are excited simultaneously. Therefore, the wave will change its polarization state at both resonance frequencies $\omega_{+}$ and $\omega_{-}$. In particular, we can expand the left-handed and right-handed circularly polarized wave (LCP and RCP) in our coordinate system with the two axes $\mathbf{J}_{-}$ and $\mathbf{J}_{+}$: $\mathbf{E}_{LCP} = \frac{1}{\sqrt{2}}\begin{pmatrix} -i \\ 1 \end{pmatrix} = a_{-}^{LCP}\mathbf{J}_{-} + a_{+}^{LCP}\mathbf{J}_{+}$, $\mathbf{E}_{RCP} = \frac{1}{\sqrt{2}}\begin{pmatrix} i \\ 1 \end{pmatrix} = a_{-}^{RCP}\mathbf{J}_{-} + a_{+}^{RCP}\mathbf{J}_{+}$. According to Eq.(8), we know that $\mathbf{J}_{\pm}$ shows strong frequency dependence, especially at $\omega_{\pm}$. Therefore, the coefficients $a_{-}^{LCP}$, $a_{+}^{LCP}$, $a_{-}^{RCP}$, $a_{+}^{RCP}$ will depend on the incident frequency. For example, at $\omega = 140 THz$, the calculated results are $a_{-}^{LCP} = -0.497 + i0.365$, $a_{+}^{LCP} = 0.365 + i0.497$, $a_{-}^{RCP} = -0.654 - i0.479$, $a_{+}^{RCP} = 0.497 - i0.654$. The transmission of LCP and RCP light can be calculated according to eq.(6) and is plotted in Figs. 3 (c) and (f). Additionally, the circular dichroism $t_{CD} = t_{RCP} - t_{LCP}$ is displayed in Fig. 3 (i). It is quite interesting that $t_{CD}$ shows a dip around the lower resonance frequency $\omega_{-}$ and a peak around the higher resonance frequency $\omega_{+}$. The reason is as follows: LCP light is composed of a larger part of $\mathbf{J}_{-}$ (which is left-handedly polarized and experiences more absorption at $\omega_{-}$ than $\omega_{+}$) than $\mathbf{J}_{+}$ (which is right-handed polarized and possesses higher absorption at $\omega_{+}$ than at $\omega_{-}$). Therefore, RCP light possesses a larger resonance absorption at $\omega_{-}$ than $\omega_{+}$: Hence $t_{RCP}(\omega_{-}) < t_{RCP}(\omega_{+})$. Reversely, LCP experiences a larger resonance absorption at $\omega_{+}$ than $\omega_{-}$, hence $t_{RCP}(\omega_{-}) > t_{RCP}(\omega_{+})$. Therefore, we can conclude that the circular dichroism satisfies

$t_{CD}(\omega_-) < t_{CD}(\omega_+)$, which explains the results in Fig.3 very well. In order to verify the above Jones Matrix formalism, we fabricated the stereo-SRR structure in Fig.1 with the method reported in [6]. The measured transmission of RCP- and LCP-light, as well as the circular dichroism are given in Fig.3 (a), (d), and (g). We also plotted the FDTD simulations results for the same structures, which are plotted in Fig. 3 (b), (e), and (h). The comparison shows that all three results agree quite well with each other. This proves that our Lagrange model is quite successful in describing the optical polarization properties of stereometamaterials.

Besides circularly polarized light, the Jones matrix formalism can also be used to calculate the polarization rotation and ellipticity of transmitted waves for any other kind of incident polarized wave. Furthermore, the Lagrange theory is not only used to calculate the optical activity in the structure given in figure 1, but can also be applied to many other metamaterials and plasmonic systems, such as spiral springs [5], gammadions [4], crosses [3] and more complex structures [2, 18].

In summary, we presented a Lagrange model to investigate the chiral optical properties of stereometamaterials. The phase retardation effect due to the 3-dimensional stacked configuration is taken into consideration. Two elliptical eigenwaves are obtained which are the basis vectors of the Jones matrix of the stereometamaterial. The polarization change for any polarized incident wave can be calculated with such a Jones matrix formalism. A stereometamaterial of twisted SRRs was fabricated and the measured circular dichrosim agrees well with our theoretical calculations. This work will stimulate many related investigations on other complex 3-D nanostructures.

This work is supported by the National Natural Science Foundation of China (No.10704036, No.10874081, No.60907009, No.10904012, No.10974090 and No. 60990320), and by the National Key Projects for Basic Researches of China (No. 2006CB921804, No. 2009CB930501 and No. 2010CB630703), as well as by DFG (FOR 557) and BMBF (3D Metamat).

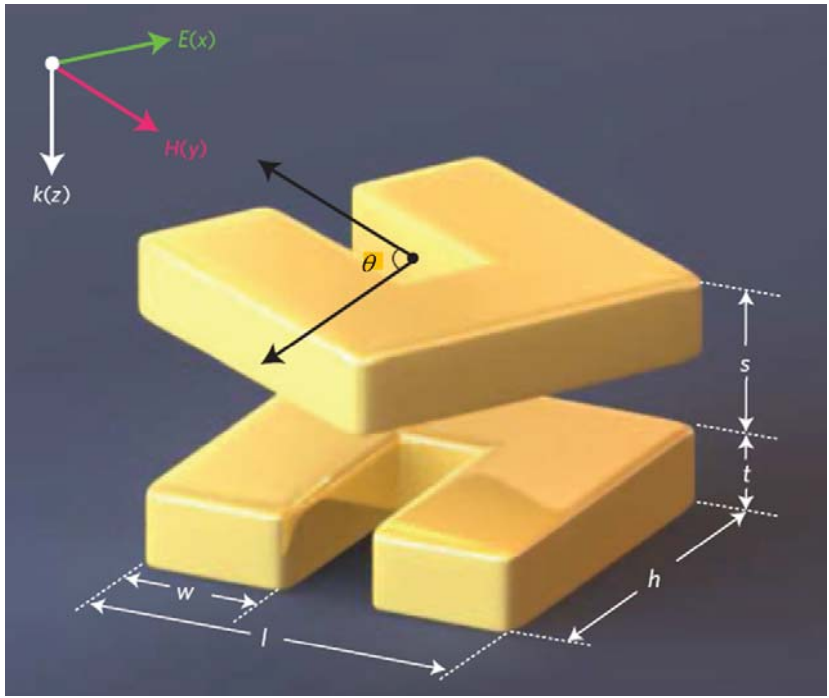

FIG.1. Schematic of the stereo-SRR dimer metamaterial with definitions of the geometrical parameters: l=230 nm, h=230 nm, w=90 nm, t=50 nm, and s=100 nm. The periods in both x and y directions are 700 nm.

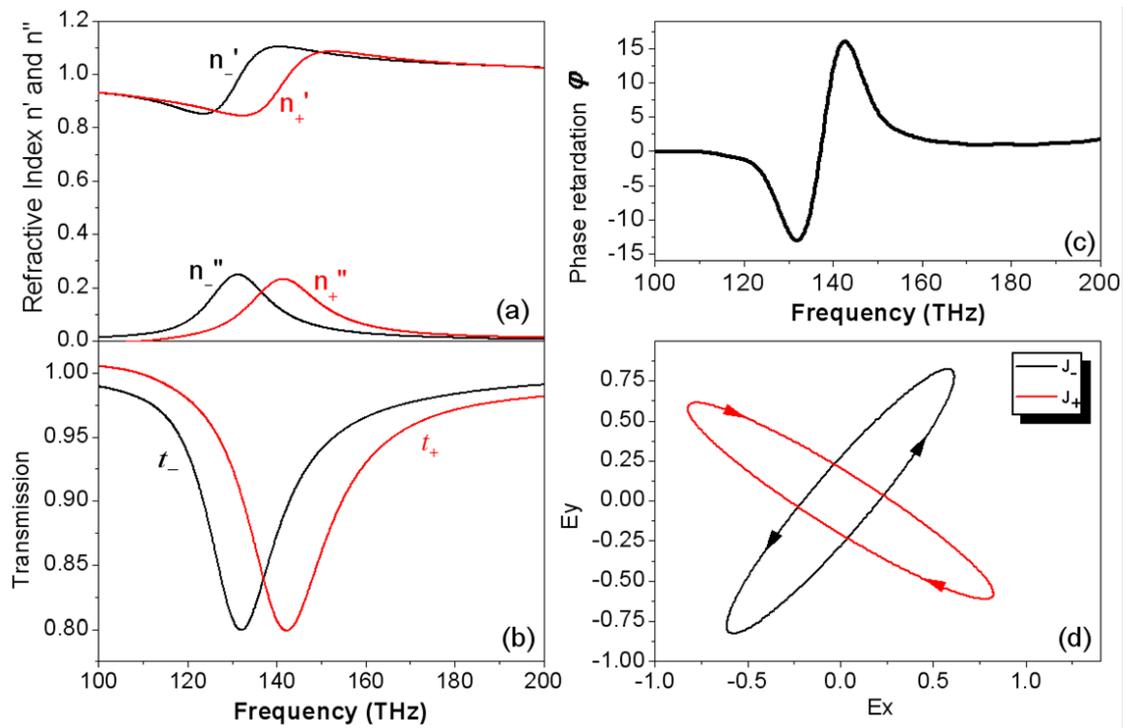

Fig. 2 (a) Refractive index of two eigenmodes in a 90° twisted SRR stereometamaterial (Lagrange model); (b) Transmission of two eigenmodes (Lagrange model); (c) Phase retardation (FDTD simulation); (d) Two elliptical polarization states of two eigenmodes at a frequency of 140 THz (Lagrange model).

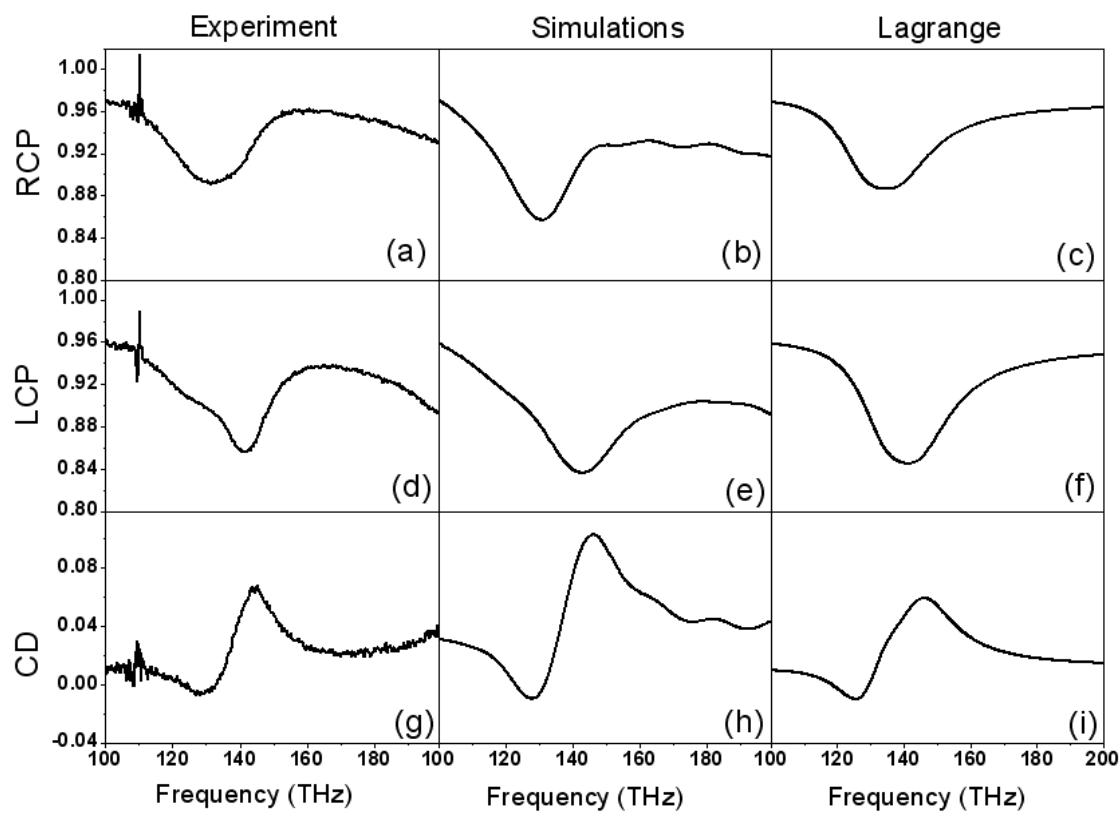

Fig.3 Transmission of left-handed (LCP) and right-handed (RCP) circularly polarized light through stereometamaterials (a-f); Circular dichroism CD (LCP-RCP) of stereometamaterials (g-i). (Experimental results: (a), (d) and (g); Simulated (FDTD) results: (b), (e), (h); Analytical Lagrange model results: (c), (f) and (i))

# Supplementary Materials for the paper

*Lagrange Model for Chiral Optical Properties of Stereometamaterials*

H. Liu[1], J. X. Cao[1], S. N. Zhu[1]

[1]*Department of Physics, National Laboratory of Solid State Microstructures, Nanjing University, Nanjing 210093 People's Republic of China*

N. Liu[2], R. Ameling[2], and H. Giessen[2]

[2]*4th Physics Institute, University of Stuttgart, Stuttgart, Germany*

In the hybridization model, each SRR can be regarded as an artificial atom; while the stereo-SRR system is like an artificial molecule. When excited by the incident EM wave, the Lagrangian of the stereo- SRR metamaterial can be rewritten as:

$$\mathcal{L} = L\left(\dot{Q}_1^2 - \omega_0^2 Q_1^2\right)/2 + L\left(\dot{Q}_2^2 - \omega_0^2 Q_2^2\right)/2 + M_m \dot{Q}_1 \dot{Q}_2 - M_e \omega_0^2 Q_1 Q_2 \cdot (\cos\theta - \alpha \cdot \cos^2\theta + \beta \cdot \cos^3\theta) \\ - \mathbf{P}_1 \cdot \mathbf{E} - \mathbf{P}_2 \cdot \mathbf{E} \cdot e^{i\varphi} \quad (1)$$

$\mathbf{P}_1 = l_{eff} \cdot \hat{\mathbf{x}} \cdot Q_1$ and $\mathbf{P}_2 = l_{eff} \cdot (\cos\theta \cdot \hat{\mathbf{x}} - \sin\theta \cdot \hat{\mathbf{y}}) \cdot Q_2$ are the induced electric dipole moments in the gap of two SRRs ($l_{eff}$ is the effective length of the single SRR), $\varphi$ is the phase retardation which arises from the distance between two stacked SRRs in the propagation direction of the EM wave. According the reference [1], the two eigenfrequencies of the stereo-SRR metamaterial are obtained as:

$$\omega_- = \omega_0 \cdot \sqrt{\frac{1+\kappa_e}{1+\kappa_m}}, \quad \omega_+ = \omega_0 \cdot \sqrt{\frac{1-\kappa_e}{1-\kappa_m}}, \quad (2)$$

where $\kappa_m = M_m/L$ and $\kappa_e = M_e \cdot (\cos\theta - \alpha \cdot \cos^2\theta + \beta \cdot \cos^3\theta)/L$ are the magnetic and electric coupling coefficients, respectively.

By introducing the Ohmic dissipation $\Re = \sigma(\dot{Q}_1^2 + \dot{Q}_2^2)/2$ and substituting Eq. (2) into the Euler-Lagrangian equations

$$(d/dt)\left(\partial\mathcal{L}/\partial\dot{Q}_i\right) - \partial\mathcal{L}/\partial Q_i = -\partial\Re/\partial\dot{Q}_i \quad (i=1,2) \quad (3)$$

we can obtain the solutions for $Q_i$ ($i=1,2$) as:

$$Q_1 = \frac{\tau \cdot (\omega^2 - i\gamma\omega - \omega_0^2) \cdot E_x + \tau \cdot (\kappa_e \omega_0^2 - \kappa_m \omega^2) \cdot (\cos\theta \cdot E_x - \sin\theta \cdot E_y) \cdot e^{i\varphi}}{(1-\kappa_m^2) \cdot (\omega^2 - i\gamma_-\omega - \omega_-^2) \cdot (\omega^2 - i\gamma_+\omega - \omega_+^2)} \quad (4.1)$$

$$Q_2 = \frac{\tau \cdot (\kappa_e \omega_0^2 - \kappa_m \omega^2) \cdot E_x + \tau \cdot (\omega^2 - i\gamma\omega - \omega_0^2) \cdot (\cos\theta \cdot E_x - \sin\theta \cdot E_y) \cdot e^{i\varphi}}{(1-\kappa_m^2) \cdot (\omega^2 - i\gamma_-\omega - \omega_-^2) \cdot (\omega^2 - i\gamma_+\omega - \omega_+^2)} \quad , \quad (4.2)$$

where $\kappa_e = M_e/L$, $\tau = l_{eff}/L$, $\gamma = \sigma/L$, $\gamma_- = \gamma/(1+\kappa_m)$ and $\gamma_+ = \gamma/(1-\kappa_m)$. Based on the material equations $\mathbf{P} = \mathbf{P}_1 + \mathbf{P}_2 = \chi\mathbf{E} = (1-\varepsilon)\mathbf{E}$ and in combination with the definition of $\mathbf{P}$ in equation (1),

the elements of the effective permittivity tensor for the stereo-SRR metamaterial can be calculated as:

$$\varepsilon_{xx} = 1 + \frac{\ell \cdot \left[\left(\omega^2 - i\gamma\omega - \omega_0^2\right)\left(1 + \cos^2\theta \cdot e^{i\varphi}\right) + \left(\kappa_e \omega_0^2 - \kappa_m \omega^2\right)\cos\theta\left(1 + e^{i\varphi}\right)\right]}{\left(1 - \kappa_m^2\right) \cdot \left(\omega^2 - i\gamma_-\omega - \omega_-^2\right) \cdot \left(\omega^2 - i\gamma_+\omega - \omega_+^2\right)}$$

$$\varepsilon_{xy} = \frac{-\ell \cdot \sin\theta \left[\left(\kappa_e \omega_0^2 - \kappa_m \omega^2\right) + \left(\omega^2 - i\gamma\omega - \omega_0^2\right)\cos\theta\right] e^{i\varphi}}{\left(1 - \kappa_m^2\right) \cdot \left(\omega^2 - i\gamma_-\omega - \omega_-^2\right) \cdot \left(\omega^2 - i\gamma_+\omega - \omega_+^2\right)}$$

$$\varepsilon_{yx} = \frac{-\ell \cdot \sin\theta \left[\left(\kappa_e \omega_0^2 - \kappa_m \omega^2\right) + e^{i\varphi}\left(\omega^2 - i\gamma\omega - \omega_0^2\right)\cos\theta\right]}{\left(1 - \kappa_m^2\right) \cdot \left(\omega^2 - i\gamma_-\omega - \omega_-^2\right) \cdot \left(\omega^2 - i\gamma_+\omega - \omega_+^2\right)}$$

$$\varepsilon_{yy} = 1 + \frac{\ell \cdot \sin^2\theta \left(\omega^2 - i\gamma\omega - \omega_0^2\right) e^{i\varphi}}{\left(1 - \kappa_m^2\right) \cdot \left(\omega^2 - i\gamma_-\omega - \omega_-^2\right) \cdot \left(\omega^2 - i\gamma_+\omega - \omega_+^2\right)}$$

(5)

Here, $\ell = l_{eff} \tau = l_{eff}^2 / L$. According to the wave equation $\mathbf{k} \times (\mathbf{k} \times \mathbf{E}) + \omega^2 \varepsilon \mu \cdot \mathbf{E} = 0$, the refractive indices of the two eigenmodes of such stereo-SRR metamaterial can be given as:

$$n_\pm = \sqrt{\left(\varepsilon_{xx} + \varepsilon_{yy} \pm \sqrt{(\varepsilon_{xx} - \varepsilon_{yy})^2 + 4\varepsilon_{xy}\varepsilon_{yx}}\right)/2}. \quad (6)$$

The corresponding polarization states of these two eigenmodes can be represented by the Jones vectors:

$$\mathbf{J}_- = \frac{1}{\mathcal{N}_-} \begin{pmatrix} -2\varepsilon_{yx} \\ (\varepsilon_{xx} - \varepsilon_{yy}) + \sqrt{(\varepsilon_{xx} - \varepsilon_{yy})^2 + 4\varepsilon_{xy}\varepsilon_{yx}} \end{pmatrix}$$

$$\mathbf{J}_+ = \frac{1}{\mathcal{N}_+} \begin{pmatrix} -2\varepsilon_{xy} \\ (\varepsilon_{xx} - \varepsilon_{yy}) - \sqrt{(\varepsilon_{xx} - \varepsilon_{yy})^2 + 4\varepsilon_{xy}\varepsilon_{yx}} \end{pmatrix},$$

(7)

where $\mathcal{N}_\pm$ are the correspondingly normalized coefficients.

For these two eigenmodes, the corresponding transmission coefficients through the stereometamaterial can be determined as [2]

$$t_\pm = \frac{4n_\pm \exp(-in_\pm kd)}{(n_\pm + 1)^2 - (n_\pm - 1)^2 \exp(-2in_\pm kd)} \quad (8)$$

Following the Jones calculus, any state in the x-y coordinate system can be transformed into a principal $\mathbf{J}_+$-$\mathbf{J}_-$ coordinate system. This Jones matrix formalism is depicted by the scheme given in figure 1. It demonstrates that our stereometamaterial can be replaced by a black box which describes an artificial polarization element with a certain Jones Matrix $\mathbf{J}_+$ and $\mathbf{J}_-$

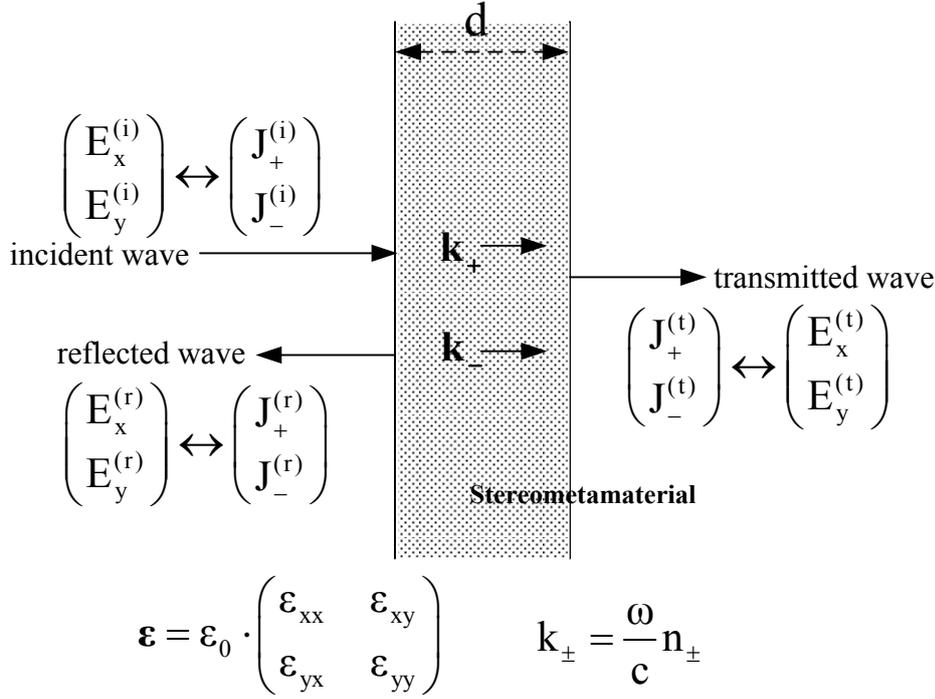

Fig. 1 Schematic for the Jones Matrix formalism in stereometamaterials.

Subsequently, the transmission coefficient for any incident wave $(E_x^i, E_y^i)$ can be calculated as

$$\begin{pmatrix} E_x^t \\ E_y^t \end{pmatrix} = \mathbf{T} \cdot \begin{pmatrix} E_x^i \\ E_y^i \end{pmatrix} = \begin{pmatrix} \hat{x} \cdot \mathbf{J}_+ & \hat{x} \cdot \mathbf{J}_- \\ \hat{y} \cdot \mathbf{J}_+ & \hat{y} \cdot \mathbf{J}_- \end{pmatrix} \begin{pmatrix} t_+ & 0 \\ 0 & t_- \end{pmatrix} \begin{pmatrix} \hat{x} \cdot \mathbf{J}_+ & \hat{y} \cdot \mathbf{J}_+ \\ \hat{x} \cdot \mathbf{J}_- & \hat{y} \cdot \mathbf{J}_- \end{pmatrix} \begin{pmatrix} E_x^i \\ E_y^i \end{pmatrix}. \quad (9)$$

Here, **T** is the Jones Matrix which is used to determine the polarization state of the transmission wave through the stereometamaterial. Eq. (9) predicts the following: When the incident EM wave is $\mathbf{J}_+ = \begin{pmatrix} \hat{x} \cdot \mathbf{J}_+ \\ \hat{y} \cdot \mathbf{J}_+ \end{pmatrix}$, the transmitted wave is $t_+ \cdot \mathbf{J}_+ = t_+ \cdot \begin{pmatrix} \hat{x} \cdot \mathbf{J}_+ \\ \hat{y} \cdot \mathbf{J}_+ \end{pmatrix}$; and when the incident EM wave is $\mathbf{J}_- = \begin{pmatrix} \hat{x} \cdot \mathbf{J}_- \\ \hat{y} \cdot \mathbf{J}_- \end{pmatrix}$, the transmitted wave becomes $t_- \cdot \mathbf{J}_- = t_- \cdot \begin{pmatrix} \hat{x} \cdot \mathbf{J}_- \\ \hat{y} \cdot \mathbf{J}_- \end{pmatrix}$. Thus, if the polarization state of the incident wave is $J_-$ or $J_+$, the transmitted wave will retain the same polarization state. For the linearly polarized incident case $\begin{pmatrix} E_0 \\ 0 \end{pmatrix}$, the transmitted wave is $\begin{pmatrix} E_0(t_+ \hat{x} \cdot \mathbf{J}_+ \hat{x} \cdot \mathbf{J}_+ + t_- \hat{x} \cdot \mathbf{J}_- \hat{x} \cdot \mathbf{J}_-) \\ E_0(t_+ \hat{y} \cdot \mathbf{J}_+ \hat{x} \cdot \mathbf{J}_+ + t_- \hat{y} \cdot \mathbf{J}_- \hat{x} \cdot \mathbf{J}_-) \end{pmatrix}$. Then, the polarization state of the transmitted wave will change.

For the cases $\theta = 0°$ and $180°$, $\varepsilon_{xy} = \varepsilon_{yx} = 0$, $\varepsilon_{yy} = 1$, $\varepsilon_{xx} \neq 1$, only the x-polarization resonance can be excited at the fundamental energy. Correspondingly, $n_+ = \sqrt{\varepsilon_{xx}}$, $n_- = 1$, $J_\pm = [1 \ 0]$. Thus, no polarization change will happen to the transmitted wave. For the case $\theta = 90°$, the elements of the permittivity tensor were obtained as:

$$\varepsilon_{xx} = 1 + \frac{\ell \cdot (\omega^2 - i\gamma\omega - \omega_0^2)}{(1-\kappa_m^2) \cdot (\omega^2 - i\gamma_-\omega - \omega_-^2) \cdot (\omega^2 - i\gamma_+\omega - \omega_+^2)}$$

$$\varepsilon_{xy} = \frac{\ell \cdot \kappa_m \omega^2 e^{i\varphi}}{(1-\kappa_m^2) \cdot (\omega^2 - i\gamma_-\omega - \omega_-^2) \cdot (\omega^2 - i\gamma_+\omega - \omega_+^2)}$$

$$\varepsilon_{yx} = \frac{\ell \cdot \kappa_m \omega^2}{(1-\kappa_m^2) \cdot (\omega^2 - i\gamma_-\omega - \omega_-^2) \cdot (\omega^2 - i\gamma_+\omega - \omega_+^2)} \quad (10)$$

$$\varepsilon_{yy} = 1 + \frac{\ell \cdot (\omega^2 - i\gamma\omega - \omega_0^2) e^{i\varphi}}{(1-\kappa_m^2) \cdot (\omega^2 - i\gamma_-\omega - \omega_-^2) \cdot (\omega^2 - i\gamma_+\omega - \omega_+^2)}$$

The refractive indices of the two eigenmodes and the corresponding polarization states are:

$$n_\pm = \sqrt{1 + \frac{(1+e^{i\varphi})\ell(\omega^2 - i\gamma\omega - \omega_0^2) \pm \ell\sqrt{(1-e^{i\varphi})^2(\omega^2 - i\gamma\omega - \omega_0^2)^2 + 4\kappa_m^2\omega^4 e^{i\varphi}}}{2(1-\kappa_m^2) \cdot (\omega^2 - i\gamma_-\omega - \omega_-^2) \cdot (\omega^2 - i\gamma_+\omega - \omega_+^2)}}$$

$$\mathbf{J}_- = \frac{1}{N_-} \begin{pmatrix} -2\kappa_m\omega^2 \\ (1-e^{i\varphi})(\omega^2 - i\gamma\omega - \omega_0^2) + \sqrt{(1-e^{i\varphi})^2(\omega^2 - i\gamma\omega - \omega_0^2)^2 + 4\kappa_m^2\omega^4 e^{i\varphi}} \end{pmatrix} \quad (11)$$

$$\mathbf{J}_+ = \frac{1}{N_+} \begin{pmatrix} -2\kappa_m\omega^2 e^{i\varphi} \\ (1-e^{i\varphi})(\omega^2 - i\gamma\omega - \omega_0^2) - \sqrt{(1-e^{i\varphi})^2(\omega^2 - i\gamma\omega - \omega_0^2)^2 + 4\kappa_m^2\omega^4 e^{i\varphi}} \end{pmatrix}$$

From Eq. (11) we know the polarization states of the eigenmodes are frequency dependent. For different frequencies, we will obtain different eigenwaves.